\documentclass[twocolumn]{revtex4}
\usepackage{psfrag}
\usepackage{amssymb}
\usepackage{indentfirst}
\usepackage[dvips]{graphicx}

\begin{document}
\title{Effects of spatial dispersion in near-field radiative heat
transfer between two parallel metallic surfaces
}
\author{Pierre-Olivier Chapuis}
\author{Sebastian Volz}
\author{Carsten Henkel*}
\author{Karl Joulain**}
\author{Jean-Jacques Greffet}
\affiliation{Laboratoire d'Energ\'etique Mol\'eculaire et
Macroscopique, Combustion\\
CNRS UPR 288, Ecole Centrale Paris\\
Grande
Voie des Vignes, F-92295 Ch\^atenay-Malabry cedex, France\\
E-mail: olivier.chapuis@em2c.ecp.fr}
\affiliation{*Universit\"at Potsdam, Institut f\"ur Physik\\
Am Neuen Palais 10,14469 Potsdam, Germany }
\affiliation{{*}*Laboratoire d'Etudes Thermiques, ENSMA\\
86961 Futuroscope Chasseneuil cedex, France}
\linespread{1}
\begin{abstract}

We study the heat transfer between two parallel metallic semi-infinite media with a gap in the nanometer-scale range. We show that the near-field radiative heat flux saturates at distances smaller than the metal skin depth when using a local dielectric constant and investigate the origin of this effect. The effect of non-local corrections is analysed using the Lindhard-Mermin and Boltzmann-Mermin models. We find that local and non-local models yield the same heat fluxes for gaps larger than $2\,{\rm nm}$. Finally, we explain the saturation observed in a recent experiment as a manifestation of the skin depth and show that heat is mainly dissipated by eddy currents in metallic bodies.
\end{abstract}

\maketitle

\section{Introduction}
Near-field radiative heat transfer has been investigated for 40 years (references \cite{Rytov,CravalhoJHT,Olivei,CarenPAA,CravalhoPAA,Hargreaves,Domoto,%
Polder,CarenIJHMT,Kuteladze,PolevoiRytov,Xu1,XuJAP,Loomis,Dorofeev,%
Pendry,KittelVacuum,PanOptics,VolokitinPRB2001,MuletAPL,Dedkov,Park,%
MuletMTE,Zhang,Zhang2,HolthausQuantum,Kittel,Bimonte}). Rytov and co-workers~\cite{Rytov} showed how to calculate thermal radiation by introducing fluctuational electrodynamics. This theory is based on the introduction of random current densities due to the thermal random motion of charges. Their correlation functions are given by the fluctuation-dissipation theorem (FDT). Cravalho, Tien and Caren~\cite{CravalhoJHT} and Olivei~\cite{Olivei} were the first to address heat transfer in the near field,
i.e., at distances smaller
than the peak wavelength $\lambda_{T}$ of the thermal radiation
spectrum.
However, they did not
consider all evanescent waves. Polder and Van Hove~\cite{Polder} were
the first to take into account all the evanescent waves by using the
formalism introduced by Rytov. They found a huge increase in the heat flux between two parallel surfaces when the gap distance $d$ becomes smaller than $\lambda_{T}$.
Rytov~\cite{PolevoiRytov} and co-workers pointed out that spatial dispersion could play a role for small gaps. Volokitin and Persson~\cite{VolokitinPRB2001} showed that spatial dispersion could be responsible for an increased heat flux in the nanometer-range by using an approximation for the non-local reflection coefficients. Loomis and Maris~\cite{Loomis} also investigated heat transfer between metallic bodies, showing the influence of the electrical resistivity. Recently, Mulet et al.~\cite{MuletAPL,MuletMTE} showed that the radiative heat transfer between dielectrics supporting surface phonon polaritons is dominated by the surface wave contribution. As a result, the heat flux is monochromatic in this case.

Several experiments have been reported. Tien's collaborators made the first measurements at cryogenic temperatures, when the near field starts at hundreds of microns. Kuteladze and Bal'tsevitch~\cite{Kuteladze} performed an analogous experiment. Hargreaves~\cite{Hargreaves} was the first one to note (at ambient temperature) an enhanced heat transfer over micrometric distances by using two parallel plates of chromium. In the end of the 1980's,
Xu~\cite{Xu1,XuJAP} et al.\ could not confirm this effect with an indium needle in front of silver. Recently, Kittel et al.~\cite{KittelVacuum,Kittel} showed a large increase in the heat exchange between a scanning probe microscope (SPM) metallic tip and a planar surface by working in the nanometer range. Surprisingly,
they also found that the increase of the heat flux levels off (saturates) at very small scales (a few $10\,{\rm nm}$). This is in striking contrast with the $1/d^3$ dependence to the distance $d$ of the density of states close to the surface. It is also in contrast with the power laws discussed by Pan et al.~\cite{PanOptics}.
This led Kittel et al.\ to suggest that the observed saturation at short distances be due to a non-local dielectric constant.
Very recently, Narayanaswamy and Chen~\cite{ArvindThesis} measured an enhancement of the heat flux
at micron distances, using dielectric polar material
and a setup similar to the one used for measurements of the Casimir force~\cite{Mohideen}.
Simultaneously, a number of groups tried to use
proximity-enhanced heat transfer to increase locally the
number of electric charge carriers.
Di Matteo et al.~\cite{DiMatteo}
reported an experimental observation in 2001.
A number of theoretical papers also present heat flux levels \cite{Raynolds,Pan,DiMatteo,Baldasaro,Whale,Chen,Reh,Laroche}. It has also been predicted that metamaterials~\cite{Zhang}, electron doping~\cite{Zhang2} or adsorbates~\cite{VolokitinPRB2004} may enhance the near-field heat transfer.

Although the enhancement of the flux becomes very large at distances on the order of a few nanometers, most of the published results use a local model of the dielectric constant. It has been pointed out that non-local effects should affect significantly the lifetime of a molecule close to a surface~\cite{Persson1982,FordWeber,StockmanNano,StockmanPRB}. This effect has also been studied in the context of the Casimir force~\cite{Sernelius,Svetovoy}. It appears to be a relatively minor
correction.
The experimental findings of Kittel et al. has revived the interest for non-local effects as the saturation observed at short distance is a very significant effect. This paper is devoted to the analysis of two questions: i) what is the origin of the saturation of the flux in the near field ? ii) what are the consequences of non-locality in the context of near-field radiative heat transfer ?

In this paper we focus on the heat flux between two parallel semi-infinite metallic substrates. We show that for a metal the s-polarized (transverse electric, TE) contribution is the leading one in the nanometric regime when using local optics. Indeed, the contribution of the familiar $1/d^2$ divergence at short distances due to p-polarized waves becomes the leading contribution only below $0.1$~nm. The saturation of the s-polarized contribution is similar to the experimental behaviour reported by Kittel~\cite{Kittel}, so that non-local corrections do not seem necessary. To further investigate this issue, we compute the near-field radiative heat transfer using two non-local models: the Lindhard-Mermin model based on the random phase approximation and its approximation in the Boltzmann-Mermin model. Both longitudinal and transverse non-local dielectric constants are included. We find that a local calculation agrees well with the non-local ones at gap distances larger than $2\,{\rm nm}$. We finally discuss the physical mechanism responsible for the saturation. We show that it is due to the magnetic fields that generate eddy currents.


\section{Near-field radiative heat flux using a local dielectric
constant}

We start the section by summarizing the derivation of the heat flux between two parallel semi-infinite bulks.  We do not consider any roughness or tilt between the surfaces.  Both semi-infinite media are assumed to be in local thermodynamic equilibrium with temperature $T_1$ and $T_2$. This allows to derive the energy radiated by random currents in medium 1 at temperature $T_1$ and absorbed in medium 2 and vice versa.
The model can be extended to inhomogeneous temperature profiles provided that the temperature variation across a distance of the order of the skin depth is negligible.
The flux per unit area is given by the normal component of the Poynting vector,
\begin{eqnarray}
    \phi =
    \langle \vec{E}(\vec{r},t) \times \vec{H}(\vec{r},t ) \rangle
    \cdot     {\bf e}_{z}
\end{eqnarray}
where the position $\vec{r}$ can be taken at the center of the gap $z = 0$ and $\langle \cdots \rangle$ denotes a statistical average. Derivations can be found in many articles~\cite{Polder,PolevoiRytov,Loomis,Dorofeev,Pendry,%
VolokitinPRB2001,MuletMTE,JoulainReview} and will not be repeated here. The final form of the heat flux is
\begin{eqnarray}
    & & \phi =
    \int_{\omega=0}^{+\infty}d\omega\,
\left[ I_{\omega}^{0}(T_{1})-I_{\omega}^{0}(T_{2}) \right]
\\
&& {} \times \sum_{\alpha \,=\, {\rm s}, {\rm p}}
\Big[
\int_{0}^{\omega / c}
\frac{KdK}{\omega^{2} / c^{2}}
\frac{
(1-\big|r_{31}^{\alpha}\big|^{2})
(1-\big|r_{32}^{\alpha}\big|^{2})}
{\big|  1-r_{31}^{s}r_{32}^{\alpha}  e^{2i \gamma_{3}d}
\big|^{2}    }
\nonumber \\
& &  +
\int_{\omega / c}^{\infty}
\frac{KdK}{\omega^{2} / c^{2}}
\frac{4\,{\rm Im}(r_{31}^{\alpha}){\rm Im}(r_{32}^{\alpha}) \,e^{-2
\gamma_{3}^{''}d}}
{\big|1-r_{31}^{\alpha}r_{32}^{\alpha}e^{-2
\gamma_{3}^{''}d}\big|^{2}}
\Big]
\nonumber
\label{eq.2}
\end{eqnarray}
where $d$ is the distance between the two interfaces,
$r_{3m}$ the reflection factor at the interface between medium $m$ and vacuum (medium $3$) for a wave with wave vector $K$ parallel to the surface and polarization $\alpha = {\rm s}, {\rm p}$. The wave vector
\begin{equation}
    \gamma_{m} = \sqrt{ \epsilon_{m} \omega^2 / c^2 - K^2 } =
    \gamma_{m}'+i\gamma_{m}''
    \label{eq:def-gamma-3}
\end{equation}
describes the propagation across medium $m$, $c$ is the speed of light,
and
\begin{equation}
    I_{\omega}^{0} = \frac{\omega^2 }{4\pi^3 c^2}
    \frac{\hbar \omega}{(e^{\hbar \omega/k_B T}-1)}
    \label{eq:def-Planck-law}
\end{equation}
is the monochromatic specific intensity of blackbody
radiation with $\hbar$, $k_{B}$ the Planck and Boltzmann constants.
We now discuss Eq.(\ref{eq.2}) that contains an integration over the ($K,\omega$) plane. This equation naturally displays
a splitting of the heat flux into s- and p-polarized waves, and into propagating ($K < \omega / c$) and evanescent waves ($K > \omega / c$). The denominators account for multiple reflections through a Fabry-P\'erot term $1 - r_{31}^{\alpha} r_{32}^{\alpha} e^{-2\gamma_{3}''d}$. The Planck function $I^{0}_{\omega}$ acts as a temperature-dependent frequency filter that cuts off frequencies much larger than $k_{B}T/\hbar$, i.e.\ beyond the near infrared at room temperature. As $\gamma_{3}'' \simeq K$ for large $K$ parallel wave vectors (deeply evanescent waves), there is also a wave vector filter ($e^{-2\gamma_{3}''d}$): wave vectors much larger than $1/2d$ do not contribute to the heat transfer at small gap sizes. This also implies that at sub-micron distances $d \ll \lambda_{T}$, the evanescent contribution is much larger than the propagating one, leading to an enhanced heat flux.


We show in Fig.2 results obtained using a local dielectric constant. We consider a non-magnetic metallic medium characterized by a Drude model $\epsilon_{1,2}( \omega ) =\epsilon_{b}-\omega_{p}^{2} / (\omega^2 + i\,\omega \nu) $ where $\epsilon_b$ accounts for the bound electron contribution, $\omega_{p}$ is the plasma frequency and $\nu$ is the damping coefficient.
This model is appropriate for frequencies up to the infrared range where the metallic response is mainly due to the conduction electrons.
In this paper, we present results either for gold ($\epsilon_b=1$, $\omega_p=1.71~10^{16}$ s$^{-1}$, $\nu=4.05~10^{13}$ s$^{-1}$) or for aluminium ($\epsilon_b=2$, $\omega_p=2.24~10^{16}$ s$^{-1}$, $\nu=1.22~10^{14 }$ s$^{-1}$, and we use in Sec. (3) $v_F=c/148$ where $v_F$ is  the Fermi velocity and $c$ is the light velocity).

Fig.2 demonstrates that the increase of the heat flux levels off below distances of 10-30 nanometers, as was found in previous papers by Polder~\cite{Polder}, Loomis~\cite{Loomis} and Volokitin~\cite{VolokitinPRB2001}. The saturation is due to a strong s-polarized contribution. Only for distances below $0.1\,{\rm nm}$ is the flux dominated by p-polarized waves, but in this regime, the local model is no longer valid (see Fig.5 below). We note that in practice, with distances in the nm-range, the s-polarized contribution dominates the heat flux.

We now discuss the behaviour of the reflection coefficients in the ($K,\omega$)-plane (see Fig.3). This points to the origin of the leading s-wave contribution. We plot the imaginary part of the reflection factors that is proportional to the heat flux [Eq.(\ref{eq.2})]. In particular, also the local density of electromagnetic states (LDOS) is controlled by the imaginary part of the reflection amplitudes, as discussed in Refs.~\cite{Pendry,Coco}.
First of all, we observe that ${\rm Im}\,r_{s}( K, \omega )$ covers a larger domain in the ($K,\omega$)-plane and takes larger values than its p-polarized counterpart.  For the latter reflection coefficient, one has at large $K$:
\begin{eqnarray}
{\rm Im}\,r_p^{31} \simeq
\frac{ \omega \nu \omega_{sp}^2 (R + 1) }{
(\omega_{sp}^2 - \omega^2)^2 + \omega^2 \nu^2 }
+ {\cal O}\left[ (\omega \sqrt{\epsilon_{1}} / c K)^2
\right]
\label{eq.4}
\end{eqnarray}
where $R = (\epsilon_{b} - \epsilon_{3}) / (\epsilon_{b} + \epsilon_{3})$ and $\omega_{sp}^2 = \omega_{p}^2 / (\epsilon_{b} + \epsilon_{3})$. If medium~3 is vacuum and the background polarization is negligible, $R = 0$, and the surface plasmon-polariton resonance occurs at $\omega_{sp} = \omega_{p} / \sqrt{2}$.  This resonance implies a peak in the near-field radiation spectrum~\cite{JoulainReview} as seen in Fig.3. It lies for typical metals in the UV, way above the frequency range that contributes significantly to the heat flux. Note that the asymptotics~(\ref{eq.4}) becomes relevant only for extremely large $K$-vectors where $K \gg \omega \sqrt{\epsilon_{1}} / c \gg \omega / c$, this is why the p-polarization becomes dominant only at very short distances (see Fig.2).
For the s-polarization, we have in the same range of $K$,
\begin{eqnarray}
{\rm Im}\,    r_s^{31} \simeq
\frac{ \omega^2 / c^2 }{ 4 K^2 }
\frac{ \omega_{p}^2 \nu }{ \omega ( \omega^2 + \nu^2 ) }
+ {\cal O}\left[ (\omega \sqrt{\epsilon_{1}} / c K)^4
\right]
\label{eq.5}
\end{eqnarray}
which tends to zero like $1/K^2$.  This is the reason why the s-polarized contribution is often discarded when looking at the asymptotic behaviour~\cite{FordWeber}.  But as shown on Fig.3(a), there is a region where ${\rm Im}(r_s)$ has large values before decaying, corresponding to the wide interval $\omega / c \ll K \ll \sqrt{|\epsilon_1|} \,\omega / c$. We detail in the Appendix the behaviour of the reflection coefficient and how to find the borders of the regions sketched in Fig.3(a). The result is an upper wave vector given by
\begin{eqnarray}
K_{\rm max} \approx \frac{\omega_p }{c  }
\label{eq.6}
\label{eq:Kmax}
\end{eqnarray}
Thus, we predict a saturation of the s-polarized heat transfer at gap distances smaller than
\begin{eqnarray}
d_{min}= \frac{ c}{\omega_p} = \frac{ \delta(\omega \gg \nu)}{\sqrt{2}}
\label{eq.7}
\end{eqnarray}
where the metal skin depth $\delta$ is defined by $1/\delta( \omega ) = (\omega/c) {\rm Im}\sqrt{ \epsilon }$.
For frequencies between
$\nu$ and $\omega_p$, $\delta \simeq
c /\omega_p$. For gold, the skin depth in this region is $\delta=\sqrt{2} c/\omega_p
\approx 25\,{\rm nm}$. It follows that the saturation distance is given by the skin depth at frequencies higher than $\nu$.
We note that for gold, $d_{min} \simeq 18\,{\rm nm}$. This is of the same order of magnitude as the cut-off distance in the experiment of Kittel et al.~\cite{Kittel}, and, incidentally, also comparable to the electron mean free path. To summarize this section, we have found that the derivation of the heat flux between two metallic surfaces using a local dielectric constant predicts a saturation of the flux at a distance given by the skin depth.

\section{Near-field radiative heat flux using a non-local model}

We now turn to a non-local description of the heat transfer. There are several reasons to investigate the role of non-local effects in the heat transfer. First of all, non-local effects becomes significant at short distances. It has been shown that non-locality can explain the anomalous skin effect~\cite{Reuter} and has a very important effect on the lifetime of an excited atom or particle near a surface~\cite{FordWeber,StockmanPRB,StockmanNano}. It has been seen that it has a significant impact in the problem of near-field friction\cite{VolokitinFriction}. It has also been suggested that saturation of the heat flux could be due to non-local effects~\cite{Kittel}.
In addition, it is desirable to analyze the interplay
between the skin depth found above and the mean free path.

Temporal dispersion (i.e., frequency dependence of optical properties) appears when the electromagnetic (EM) field varies on a timescale comparable to the microscopic timescales of the medium where it propagates. A non-local behaviour (i.e. spatial dispersion or k-dependence of the optical properties) is expected if the EM
field varies appreciably on length scales given by the microscopic structure of the medium.

For metals, there are several microscopic length scales related to the Fermi velocity $v_{F}$ of the conduction electrons.
A first one is the electron mean free path $v_F/\nu$, typically $20\,{\rm nm}$ for gold at ambient temperature in the bulk.
The second one is the charge screening length in a plasma of electrons called the Thomas-Fermi length, on the order of $v_F/\omega_p$. A third length is the Fermi wavelength $1/k_F = \hbar / m_{*} v_{F}$ (where $m_{*}$ is the effective mass of the electron). It sets a lower limit for the spatial variations of the electron density in the metal and is often comparable to the Thomas-Fermi length. 
The fourth characteristic length is the distance $v_{F} / \omega$ travelled by an electron during one period of an applied EM field.
This length governs an enhanced absorption by evanescent waves with $K > \omega / v_{F}$. This process is called Landau damping and consists in the creation of electron-hole pairs by absorption of photons.

In order to account for the bulk effects we use two different dielectric functions: the Lindhard-Mermin (LM) and the Boltzmann-Mermin (BM) formulae~\cite{FordWeber}. The LM dielectric function is also known, e.g., as the random phase approximation (RPA)~\cite{FordWeber}, Kliewer-Fuchs~\cite{KliewerFuchs} constants or jellium ones. Other types of non-local dielectric functions are possible: the hydrodynamic model is an approximation at small wave number~\cite{Halevi};  Feibelman's model~\cite{Feibelman} focusses on surface effects and has difficulties in taking bulk absorption into account, which is playing a significant role in heat transfer. We follow the notations of Ford and Weber for the longitudinal and transverse dielectric functions~\cite{FordWeber}
\begin{eqnarray}
\epsilon_{l}^{LM}(k,\omega)  =
\epsilon_{b}+\frac{3\omega_p^2}{\left( \omega+ i
\nu \right)}\frac{u^2~f_{l}(z,u)}{\left[ \omega+i\,\nu \frac{f_{l}(z,u)}{f_{l}(z,0)} \right]}
\label{eq.8}
\end{eqnarray}
\begin{eqnarray}
&& \epsilon_{t}^{LM}(k,\omega)  =
\epsilon_{b}-\frac{\omega_{p}^{2}}{\omega^{2}(\omega+i \nu)} \{
\omega \big[f_{t}(z,u)
\label{eq.9}\\
&& \quad
{}-3z^{2}f_{l}(z,u)\big]
+ i\,\nu \left[f_{t}(z,0)-3z^{2}f_{l}(z,0)\right] \}
\nonumber
\end{eqnarray}
where $\epsilon_{b}$ is the bulk contribution to the dielectric constant. It describes the interband contributions and it is constant
in the following as these transitions do not play any role in the frequency range that we address. The Lindhard functions $f_{l,t}(z, u)$ have arguments $z=k/2k_{F}$ and $u=(\omega + i\,\nu)/kv_{F}$ with $k_{F}$ the Fermi wave vector, and are given by
\begin{eqnarray}
\nonumber
f_{l}(z,u) & = &
\frac{1}{2}+\frac{1-(z-u)^2}{8z}~\,\ln\frac{z-u+1}{z-u-1}\\
& &+\frac{1-(z+u)^2}{8z}~\,\ln\frac{z+u+1}{z+u-1}
\label{eq.10}
\end{eqnarray}
\begin{eqnarray}
\nonumber
f_{t}(z,u)  =
\frac{3}{8}(z^2+3u^2+1)-3\frac{\left[1-(z-u)^2\right]^2}{32z}\\
~\,\ln\frac{z-u+1}{z-u-1}
-3\frac{\left[1-(z+u)^2\right]}{32z}~\,\ln\frac{z+u+1}{z+u-1}
\label{eq.11}
\end{eqnarray}
The limit $u \to 0$ has to be taken with a positive imaginary part
so that
\begin{eqnarray}
f_{l}(z,0)=\frac{1}{2}+\frac{1-z^2}{4z} \,\ln | \frac{z+1}{z-1} |
\label{eq.12}
\end{eqnarray}
and
\begin{eqnarray}
f_{t}(z,0)=\frac{3}{8}(z^2+1)-3\frac{(1-z^2)^2}{16z} \,\ln |
\frac{z+1}{z-1} |
\label{eq.13}
\end{eqnarray}
A semiclassical approximation of these formulae is obtained for
wave vectors $k$ much smaller than $k_{F}$, taking $z=0$. This
gives the Boltzmann-Mermin formulas
\begin{eqnarray}
\epsilon_{l}^{BM}(k,\omega)=\epsilon_{b}+\frac{3\omega_p^2}{\left( \omega+
i\,\nu \right) }\frac{u^2~f_{l}(0,u)}{\left[ \omega+i\,\nu f_{l}(0,u) \right]}
\label{eq.14}
\end{eqnarray}
\begin{eqnarray}
\epsilon_{t}^{BM}(k,\omega)=\epsilon_{b} -
\frac{\omega_{p}^{2}}{\omega^{2}(\omega + i\,\nu)} f_{t}(0,u)
\label{eq.15}
\end{eqnarray}
where
\begin{eqnarray}
f_{l}(0,u)=1-\frac{u}{2}\,\ln \frac{u+1}{u-1}
\label{eq.16}
\end{eqnarray}
and
\begin{eqnarray}
f_{t}(0,u)=\frac{3}{2}u^{2}-\frac{3}{4}u(u^2-1)\,\ln \frac{u+1}{u-1}
\label{eq.17}
\end{eqnarray}
A few remarks are in order here. First, the Drude formula is recovered at small $k$ ($u$ large and $z$ small).
Second, the variable $u$ compares $k$ to a combination of the mean free path $v_{F}/\nu$ and the distance covered by an electron during a period of the field $v_{F}/\omega$, that can be considered as an ``effective mean free path"~\cite{Carsten_NonLocal}. Third, at very large wave vectors, the logarithms in Eqs.(\ref{eq.16},\ref{eq.17}) describe Landau damping. Indeed, even for $\nu=0$, they imply ${\rm Im}(\epsilon)>0$ for $k>\omega/v_F$~\cite{StockmanPRB}. Finally, it is seen that at very large wave vectors, there is a sharp cut-off in the imaginary parts of the Lindhard-Mermin dielectric functions:
\begin{eqnarray}
\epsilon_{t}^{LM}(k \gg k_{F})=\epsilon_{b} +
\frac{8}{5}\frac{\omega_p^2}{\omega^2}
\frac{k_{F}^2}{k^2} + i\,\nu\frac{4}{\omega}
\frac{\omega_p^2}{v_{F}^2}\frac{k_{F}^2}{k^4}
\label{eq.18}
\end{eqnarray}
\begin{eqnarray}
\epsilon_{l}^{LM}(k\gg k_{F})=\epsilon_{b}+\frac{4
\omega_p^2}{v_{F}^2}\frac{k_F^2}{k^4}+i\,\nu \frac{16\omega \omega_p^2}{v_{F}^4}
\frac{ k_F^4}{k^8}.
\label{eq:eps-l}
\end{eqnarray}
Thus, fields oscillating with spatial periods smaller than half the Fermi wavelength cannot be screened by the electron plasma.

We now account for microscopic surface effects that modify the reflection amplitudes. For the sake of simplicity, we use the infinite barrier model (also known as SCIB) that considers that electrons undergo specular reflection at the boundary~\cite{FordWeber}. A model considering diffuse reflection of electrons is also available~\cite{Nazarov}. In our specular case, the reflection coefficients are computed in terms of surface impedances as follows
\begin{eqnarray}
r_p^{31} = \frac{\gamma_3/(\omega \epsilon_{3}) - Z_{p}}{\gamma_3/(\omega
\epsilon_{3}) + Z_{p}}
\label{eq.19}
\end{eqnarray}
\begin{eqnarray}
r_s^{31} =
\frac{Z_{s} - \omega/(c^2\gamma_{3})}{Z_{s} + \omega/(c^2\gamma_{3})}
\label{eq.20}
\end{eqnarray}
with
\begin{eqnarray}
    && Z_{s}(K,\omega)
     =
     \frac{1}{c}\frac{\left(\vec{z} \times
     \vec{K}\right).{\vec{E_{1}}}}{\vec{K}.\vec{B_{1}}}\\ \nonumber
     &&
     =\frac{2i}{\pi \omega}
\int\limits_{0}^{\infty}{dq_{z}\frac{1}{\epsilon_t(k,\omega)-(ck/\omega)^2}}
\label{eq.21}\\
& &Z_{p}(K,\omega)
 = \frac{-1}{c} \frac{\vec{K}.\vec{E_{1}}}{\left(\vec{z}
 \times \vec{K}\right).\vec{B_{1}}}\\ \nonumber
 &  &
    =
    \label{eq.22}\\
&&    \frac{2i}{\pi \omega}
\int{\frac{dq_{z}}{k^2}\left(
\frac{q_z^2}{\epsilon_t(k,\omega)-(ck/\omega)^2}
+\frac{K^2}{\epsilon_l(k,\omega)}\right)}
\nonumber
\end{eqnarray}
where under the integral,
$k^2 = K^2+q_z^2$.  $\vec{K}$ is the unit vector in the direction  of the parallel wave vector $K$. As we account for spatial dispersion by using a non-local model, the reflection coefficients depend on $\omega$ and $K$ in a more complicated way than the Fresnel formulas.

One should note that in this approach, we do not tackle several effects that occur on the atomic (sub-nm) scale. The electron
density, which is modified near the interface, is treated here with a step form and the addition of surface currents~\cite{KliewerFuchs,FordWeber}.
Several authors~\cite{Feibelman,FordWeber} showed that a self-consistent
calculation leads to a continuous variation of the electron density between the bulk density and vacuum and that this can be described by an effective mean displacement of the surface, of the order of a few angstr\"oms.
Phenomena like electron tunneling also occur as the two surface approach each other on this scale and mutually influence their electron density profiles. We do not take this tunneling into account as it is clearly negligible in the nanometer range.

We show on Fig. 4(a) the imaginary part of $r_{p}$, at fixed $\omega$. It is related to the local LDOS (see section IV).  An interesting finding is that the local description leads to a plateau for large $K$ (non-retarded approximation) that
does not agree for any value of $K$ with the non-local model. The local quasistatic approximation that has been often used, thus yields an incorrect value of ${\rm Im}(r_p)$ for a very broad range of frequencies. The curve labelled `longitudinal quasistatic' is based on neglecting the first term in Eq.(\ref{eq.22}), involving the transverse part of the dielectric function. We see that this term nevertheless contributes at wave vectors $K < 1/\delta(\omega) \approx 10^8\,{\rm m}^{-1}$. For larger $K$, the non-local calculation leads to an increase of ${\rm Im}\,r_{p}$ by roughly one order of magnitude, that we attribute to Landau damping.
Finally, we observe that for wave vectors larger than $k_{F} \approx 10^{10}\,{\rm m}^{-1}$, the non-local models predict a strong decay of ${\rm Im}(r_p)$ as compared to the local model.

The s-polarized reflection coefficient ${\rm Im}(r_s)$ is plotted in Figure 4(b). Differences to the local calculation are barely visible
in the domain
$K<5~10^8\,{\rm m}^{-1}$ where ${\rm Im}(r_s)$ takes significant values and contributes to the heat transfer.
We thus expect only small corrections to heat transfer from the non-local models.


Figure 5 presents the heat flux as a function of the gap distance. We display the fluxes due to s and p polarizations when using both a local and the two non-local models introduced above. Although the validity of the models is questionable for distances smaller than $1\,{\rm nm}$, we display the flux at smaller distances in order to analyze their physical content when $d \rightarrow 0$. What is important here is that the local and non-local heat fluxes are identical up to distances on the order of the Thomas-Fermi length $v_{F} / \omega_{p}$. It appears that the small modifications of ${\rm Im}(r_s)$ give the same final result after integration over $K$ and $\omega$.
A small increase of the heat flux~\cite{VolokitinPRB2001} due to the onset of Landau damping is observed in the p-polarized contribution, but in a regime where s-waves dominate and level off. Another observation is that the two non-local models are superimposed, showing that the Thomas-Fermi length is sufficient to describe the large $K$ decay of the dielectric constant. Finally, at very short distances (below the Thomas-Fermi length or the Fermi wavelength), the non-local models remove the $1/d^2$ regime of the p-polarized flux.

We now illustrate how the non-local models suppress this $1/d^2$ dependence. We have plotted in Fig.6 the p-polarized contribution to the heat flux in the $(K, \omega)$-plane, but removing the decay term $e^{-2{\rm Im}(\gamma_{3})}$ and the Planck function $I_{\omega}^{0}(T)$ that act as filters. What we plot is thus ${\rm Im}(r_p^{31})^2 / | 1 - (r_p^{31})^2e^{-2 \gamma_3'' d}|^2$. Figs.6 show a locus that follows the dispersion relation of the surface plasmon-polariton. It is seen that it has two branches~\cite{Sernelius,Economou69}. They split at a wave vector of order $1/d$ that is pushed towards large $K$ as the gap size is decreased. When non-locality is included, the flat asymptote at frequency $\omega_{sp} = \omega_p/\sqrt{2}$ for large values of $K$ becomes dispersive and approaches $\omega = v_{F} K$ in Fig.6(b). But what is important here is that the far IR branche of the resonance will not be able to be shifted to the large $K$ region when the gap size decreases because of the cut-off at $\omega = v_{F} K$. This removes the divergence of the heat flux due to the p-polarized evanescent contribution in Eq.\ref{eq.2} when $d \to 0$.
It provides an intrinsic cutoff at large $K$ which is different from the distance $d$.


The main conclusion of this section is that the local calculation is in practice sufficient when computing heat fluxes between two metallic surfaces a few nanometers apart. The second conclusion is that
non-locality removes the universal heat flux divergence at short distance as expected.


\section{Discussion and concluding remarks}

In this last section, we try to gain some insight on the physical mechanisms responsible for the near-field heat transfer in s-polarization between two parallel interfaces.  We have plotted in Figure 7 the LDOS~\cite{Joulain2003,JoulainReview} near a metallic/vacuum interface in vacuum. We recall that the local density of energy is the product of the LDOS by the mean energy of an oscillator given by $\hbar \omega/(e^{\frac{\hbar \omega}{k_B
T}}-1)$.
The LDOS is split in four contributions: magnetic and electric fields, s- and p-polarization. For instance, the contribution of the evanescent s-%
polarized magnetic field to the LDOS is given by
\begin{equation}
\rho^{M}_{s}(z,\omega) = \rho_{v}\int_{\omega/c}^{+\infty}
\frac{dK}{2|\gamma_{3}|}\frac{cK}{\omega}f(K,\omega){\rm
Im}(r_{s})~e^{-2\gamma_{3}''z}
\end{equation}
where $\rho_{v}(\omega)=\omega^{2}/\pi^{2}c^{3}$ is the vacuum density of states and $f(K,\omega)=2(\frac{cK}{\omega})^{2}-1$. Again, the properties of the material control the LDOS via ${\rm Im}(r_s)$.
Fig.7 shows that the propagating terms are negligible. Furthermore, the leading contribution in the infrared ($\omega \approx 10^{13} \ldots 10^{15}\,{\rm s}^{-1}$, where the room-temperature thermal spectrum peaks) is clearly due to s-polarized magnetic fields. It follows that a metallic half-space generates a very large magnetic energy in a vacuum close to the surface. This quantity is relevant to analyze the heat transfer through an interface. Indeed, as the magnetic field is continuous through an interface with a non-magnetic material, the magnetic field penetrates without reflection.

The large value of the magnetic density of energy due to s-polarized waves near a metallic interface has been discussed recently~\cite{Joulain2003,Coco}. Whereas the ratio $c\vert \vec{B}\vert/\vert \vec{E}\vert$ takes a fixed value of $1$ for propagating waves, it becomes frequency-dependent for evanescent waves ($K/k_0 >1$).
For s-polarized evanescent waves, one can show using the Maxwell-Faraday equation that this ratio is given by$\sqrt{f(K,\omega)}\simeq \sqrt{2}\, K/k_0$.  Magnetic fields dominate in s-polarization. For p-polarized waves, the opposite trend $\vert \vec{E}\vert/c\vert \vec{B}\vert \simeq \sqrt{f(K,\omega)}$ is found, showing that electric fields dominate. If we want to know which of the magnetic s-polarized waves or the electric p-polarized waves give the leading contribution to the LDOS, we have to compare the products $f(K,\omega){\rm Im}(r_{s})$ and $f(K,\omega){\rm Im}(r_{p})$. As we have seen, the s-polarized reflection coefficient is larger than than ${\rm Im}(r_p)$ for a metal at infrared frequencies and below, so that finally, the LDOS is dominated by its s-polarized magnetic component as seen on Fig.7.

 It follows that retardation plays a key role as observed in Ref. \cite{VolokitinPRB2001}. Accordingly, the heat transfer between a metallic nanoparticle and a half space \cite{Pendry,VolokitinPRB2001,MuletAPL} must be revisited accounting for magnetic energy. It will be shown that the magnetic dipole yields the leading contribution \cite{Coco}.

The large magnetic fields can be traced back to the current density in the material. In s-polarization, the electric field $\vec{E}$ is tangential to the metallic interface and therefore continuous. It drives a surface current flowing within the skin depth $\delta$, with an amplitude roughly given by $\sigma
E$. This suggests the following mechanism for the heat transfer between metallic surfaces: fluctuating currents flowing parallel to the interface within the skin depth in medium~1 generate large magnetic fields at IR frequencies. These
fields penetrate into medium~2 and generate large eddy currents which are dissipated by the Joule effect. In other words, radiative heat transfer in the near field is similar to nanoscale induction heating at infrared frequencies.


We have seen in section~2 that the skin depth plays a key role~\cite{CarstenOpticsCom}. The above argument provides a simple picture for the phenomenon. The skin depth depends on the frequency.  We stress that the cutoff distance seen by Kittel~\cite{Kittel} and that we found above is linked to the skin depth evaluated at the frequencies contributing to the largest parallel wave vectors, $\omega \b nu$. For gold, this skin depth
is $\delta=\sqrt{2} c/\omega_p
\approx 25\,{\rm nm}$.
Our analysis leads to a number of predictions that should be measurable. Measurements of  the heat transfer such as reported by Kittel should be able to detect the skin depth dependence by changing the metals.
As seen on Figure 8, the plasma frequencies of a number of metals are not very different. They all give (local) cut-off distances in the range of 10 to 200\,nm. The differences should be measurable. A material like cobalt is expected to saturate at larger distances than metals like copper, gold or aluminium. Interestingly, cobalt could also be a test-case study for the saturation due to non-locality as the p-polarized contribution becomes larger than the s-polarized contribution near $1\,{\rm nm}$. Another interesting issue is the heat flux between two different metals. We expect a saturation distance governed by the smallest skin depth due to the product ${\rm Im}(r_s^{31}) {\rm Im}(r_{s}^{32})$ in the heat flux formula.

To summarize, we have shown that the radiative heat flux between two parallel metallic surfaces saturates when the gap size reaches a distance equal to the skin depth at a frequency equal to $\nu$. We have shown that the leading contribution to the flux is due to eddy currents generated in the medium. The non-local effects have been studied. They do not significantly affect the s-polarized fields but introduce a cut-off in the $K$ dependence of the p-polarized fields. This cut-off removes the $1/d^2$ dependence of the flux at short distances. As the s-polarized fields dominate the heat transfer between metallic surfaces, the non-local corrections are negligible. Finally, we observed that the cut-off distances seem to be in the range of 10 to 200\,nm for many metals.

\section*{APPENDIX}
We explain in this section how we estimate the limits of the domain in the $(K,\omega)$ plane where ${\rm Im}(r_s)$ contributes to the heat flux. As is shown on Figure 3, the $(K,\omega)$ plane can be divided into four areas. Point A is the intersection of the 4 borders.  In all the cases, we consider only evanescent waves: $K \gg k_0$ with $k_0=\omega/c$.

We address first the division of the $(K,\omega)$ plane between large $K$ and smaller values. This underlines the different behaviours of regions 1 and 3 on one hand, and regions 2 and 4 on the other hand. The perpendicular wave vector $\gamma_{1}$ is given by
\begin{eqnarray}
K^2+\gamma_{1}^2 = \epsilon_1 k_{0}^2
\label{eq.28}
\end{eqnarray}
where $k_{0} = \omega / c$.
This shows that we have two regimes. To leading order, we have $\gamma_{1}^2 \simeq -K^2$ at very large $K$ (regions 2 and 4) and $\gamma_{1}^2 \simeq \epsilon_1 k_0^2 $ at smaller $K$ (region 1 and 3). The transition occurs at a critical wave vector $K^2 \simeq |\epsilon_1  k_0^{2}|$. This gives a critical wave vector given by
\begin{eqnarray}
K_c(\omega) = \sqrt{|\epsilon_1(\omega)|}k_0 \approx
\frac{ \omega_{p} }{ c }
\sqrt{ \frac{ \omega }{ |\omega + i \, \nu| } }
\label{eq.29}
\end{eqnarray}
where the last equality applies to the Drude model at frequencies $\omega \ll \omega_{p}/ \sqrt{\epsilon_b}$. Values of $r_{s}$ in both regimes are now given. To leading order,
one finds
\begin{equation}
    r_{s} \approx \left\{
    \begin{array}{ll}
	\displaystyle
           -1-2 \frac{i K}{\sqrt{\epsilon_{1}}\,k_{0}}

	& \mbox{(region 1,3)}
		\\[1ex]
	\displaystyle
	\frac{k_{0}^2}{4K^2}( \epsilon_{1} - 1 )
	& \mbox{(region 2,4)}
	
	\end{array} \right.
    \label{eq:rs-asymptotics}
\end{equation}
At large $K$, ${\rm Im}(r_s)$ decreases to small values that do not
contribute significantly to the heat flux integral.

We address now the horizontal division of Fig.3.
The upper region is given by domains 3 and 4 and the lower one by
domains 1 and 2.  This limit is due to the different behaviours of
$\epsilon(\omega)$ if $\omega \ll \nu$ (domains 1 and 2) or $\omega
\gg \nu$ (domains 3 and 4).  The first two asymptotic orders are
\begin{equation}
    \epsilon_{1}( \omega ) \approx
    \left\{
    \begin{array}{ll}
	\displaystyle
	i \omega_p^2/\omega \nu - \frac{\omega_p^2}{\nu^2}
	& \mbox{(region 1,2)}
	\\[1ex]
	\displaystyle
	- \frac{ \omega_p^2 }{\omega^2}
	+ i \frac{ \omega_p^2 \nu}{ \omega^3 }
	& \mbox{(region 3,4)}
	\end{array} \right.
    \label{eq:eps-asymptotics}
\end{equation}
The low-frequency expression is also known as the Hagen-Rubens formula.
In Table~\ref{t:rs-asymptotics}, we give the corresponding asymptotics
for ${\rm Im}\,r_{s}$ in the four regions.

\begin{linespread}{1.8}
\begin{table}[bht]
\center
\begin{tabular}{|c|c|c|}
\hline
Region & Characteristics & ${\rm Im}(r_s)$       \\
\hline
\hline
1&far IR, small $K$ &
$
\frac{\sqrt{2\nu}c}{\omega_p}\frac{K}{\sqrt{\omega}}$\\
\hline
2&far IR, large $K$&$\frac{\omega_p^2}{4\nu c^2}\frac{\omega}{K^2}$\\
\hline
3&near IR, small $K$&$ \frac{\nu c}{\omega_p}\frac{K}{\omega}$\\
\hline
4&near IR, large $K$&$\frac{\omega_p^2 \nu}{c^2}\frac{1}{\omega K^2}$
\\
\hline
\end{tabular}
\caption{Asymptotic behaviour of ${\rm Im}(r_s)$. A local Drude model is taken for $\epsilon( \omega )$ with plasma frequency $\omega_{p}$ and relaxation rate $\nu$.}
\label{t:rs-asymptotics}
\end{table}
\end{linespread}

As a function of frequency, the critical wave vector behaves like $K_{c} \approx (\omega_{p} / c) (\omega / \nu)^{1/2}$ in the far infrared (small frequencies) and like $K_{c} \approx \omega_{p} / c$ for larger frequencies. These two lines cross at $\omega \approx \nu$ which is the point A marked in Fig.3. At this point, the imaginary part of $r_{s}( K, \omega )$ reaches its maximum.

According to Eq.(\ref{eq.29}), ${\rm Im(r_s)}$ takes significant values for $K$ lower than $K_{c}=\omega_p / c$. This limit yields a saturation length $1/K_{c}=c/ \omega_p$. Note that this length is related to the skin depth as $\delta=\frac{1}{{\rm Im}(\sqrt{\epsilon_1 k_0})} \simeq \frac{c}{\sqrt{2} \omega_p}$. At low frequencies (regions~1 and~2), $\epsilon$ is purely imaginary, leading to $\delta \simeq \sqrt{2} / K_{c}$, while in the high frequency regions~3 and~4, $\delta \simeq 1 / K_{c}$. Hence, at each frequency, the cutoff wave vector is essentially given by the inverse skin depth.

\section*{Acknowledgments}
We thank M. Laroche, M.I. Stockman and V.B. Svetovoy for useful discussions. We
acknowledge the support of the Agence Nationale de la Recherche under contract ANR06-NANO-062-04
{
\small
\linespread{1}

}
 \end{document}